\documentclass[preprint,aps]{revtex4}
\usepackage{graphicx} % Include figure files
\usepackage{dcolumn} % Align table columns on decimal point
\usepackage{bm} % bold math
\usepackage{amsmath,amssymb}
\begin{document}

\title{Suitability of carbon nano\-tubes grown by chemical vapor deposition for electrical devices}

\author{B.\ Babi{\'c}, J.\ Furer, M.\ Iqbal and C. Sch{\"o}nenberger}
\affiliation{Institut f\"ur Physik, Universit\"at Basel,
Klingelbergstr.~82, CH-4056 Basel, Switzerland}

\begin{abstract}
Using carbon nano\-tubes (CNTs) produced by chemical vapor deposition,
we have explored different strategies for the preparation of
carbon nanotube devices suited for electrical and mechanical measurements.
Though the target device is a single small diameter CNT, there is
compelling evidence for bundling, both
for CNTs grown over structured slits and on rigid supports.
Whereas the bundling is substantial in the former case,
individual single-wall CNTs (SWNTs) can be found in the latter.
Our evidence stems from mechanical and electrical measurements on contacted tubes.
Furthermore, we report on the fabrication of low-ohmic contacts to SWNTs. We compare
Au, Ti and Pd contacts and find that Pd yields the best results.
\end{abstract}

\maketitle

% ------------------------------------------------------------------
% Main text
% ------------------------------------------------------------------

The present work is structured in two main sections. The first is
devoted to our results on
carbon nano\-tubes (CNTs) grown by chemical vapor deposition (CVD)
emphasizing on the problem of CNT bundling, which
occurs during growth. The second section discusses our results on the contacting of
CVD-grown tubes using the metals Au, Ti and Pd.

\section{Supported and suspended carbon nano\-tubes prepared by CVD}

The full control and understanding of structural
and electronic properties of carbon nano\-tubes remain a major challenge towards their
applications in nanoelectronics. Today, there exists several
different production methods of carbon nano\-tubes (CNTs). Among
them, chemical vapor deposition (CVD) emerged
\cite{growth1,Hafner1,Dai1} as the most prominent one for the
investigation of the electronic and electromechanical
properties of CNTs. The most important advantages of the CVD method are
that CNTs can be grown at specific locations on the substrate and
at lower temperatures with simpler equipments as compared to the arc
discharge and laser ablation methods. However, CNTs grown with
this method vary in a quality and display a rather large
dispersion in diameter which might
be a sever problem for potential applications. Following the
published recipes, we found that CVD grown CNTs differ
dramatically if they are grown supported on a substrate or
suspended over structured slits. This suggests that
the nanotube-substrate interaction plays
an important role in the final product in addition
to growth parameters and catalysts.

\subsection{Growth method}

Two types of catalysts are used for the growth of CNTs. The first
catalyst, which we will name catalyst 1, is similar to that
described in Ref.~\cite{Hafner1}. The catalyst suspension consists
of \mbox{$1$\,mg} iron nitrate seeds
(\mbox{Fe(NO$_3$)$_3\cdot$9H$_2$O}) dissolved in \mbox{10\,ml} of
isopropanol. The other catalyst, which we will call in the rest of
the paper catalyst 2, has been prepared similar to that described
in Ref.~\cite{Dai1}. To \mbox{15\,ml} of methanol, \mbox{15\,mg}
alumina oxide, \mbox{20\,mg} \mbox{Fe(NO$_3$)$_3\cdot$9H$_2$O} and
\mbox{5\,mg} \mbox{MoO$_2$(acac)$_2$} are added. Both suspensions
are sonicated for \mbox{1\,hour}, stirred overnight and sonicated
every time for at least \mbox{20\,min} before deposition on the
substrate \cite{substrate}. A drop of the suspension is placed on
a bare substrate surface or on a substrate with predefined
structured areas by electron-beam lithography (EBL) or optical
lithography in the corresponding resist. After spinning at
\mbox{2000\,r.p.m} for \mbox{40\,sec}, the substrate is baked at
\mbox{150\,$^\circ$C} for \mbox{5\,min}, followed by lift-off. The
CVD growth of CNTs is performed in a quartz-tube furnace between
\mbox{$750-1000$\,$^\circ$C} at atmospheric pressure using
different gases. For catalyst 1 we used a mixture of either ethylene
or methane with hydrogen and argon with respective flow rates of
\mbox{$2$}, \mbox{$400$}, and \mbox{$600$\,cm$^3$/min}~\cite{flowmeters}.
For the catalyst 2, we have used a mixture of methane and argon with
respective flow rates of \mbox{$5000$} and
\mbox{$1000$\,cm$^3$/min}~\cite{flowmeters}.
During heating and cooling of the
furnace, the quartz tube is continuously flashed with argon to
reduce the contamination of the CNTs and to avoid burning them
once they are produced.

\subsection{Results and Discussion}

Carbon nano\-tubes which are grown at the same temperature but with the two mentioned catalysts on
thermally oxidized silicon substrates show similar characteristics.
In both cases there is a profound temperature dependence. At relatively
low temperatures (\mbox{750-850\,$^\circ$C}) predominantly
individual MWNTs or ropes of SWNTs are obtained with high yield.
At intermediate temperatures (\mbox{850-975\,$^\circ$C})
individual SWNTs are grown with a typical diameter of \mbox{2\,nm}
or thin bundles of SWNTs, but with less yield than at lower
temperatures. At high temperatures (\mbox{$>$1000\,$^\circ$C}),
the substrate and the CNTs are often found to be covered with an
additional material, which is most likely amorphous
carbon. Carbon nano\-tubes used in transport measurements have been
solely produced at the intermediate temperature range. Fig.~1a shows
a scanning electron microscope (SEM) image of CNTs grown from
\mbox{catalyst~1} on a \mbox{Si/SiO$_2$} substrate.

For the purpose of mechanical and electromechanical studies,
CNTs have been grown over structured slits
patterned in \mbox{Si$_3$N$_4$}, an example of the outcome
is shown in Fig.~1b. It is expected that for
sufficiently long CNTs thermal vibrations should be readily
observed with transmission and scanning-electron microscopy (TEM and SEM) \cite{Ebbesen1,Heer,us2}.
This holds only, however, for `small' diameter tubes, because the
vibration amplitude is strongly reduced with increasing diameter $d$
according to (\mbox{$\sim1/d^2$}). Only individual SWNTs are
expected to show a substantial vibration amplitude which could be observed in
SEM. We suggest this as a simple check to distinguish individual from
bundled SWNTs. Fig.~1b shows a representing SEM image of suspended CNTs spanning over long
distances ({$L>$1\,$\mu$m}). None of the visible `strings' display observable
vibrations. This is not surprising considering the observed
CNT branches.
Clearly, in this case the CNTs must be bundled.
This bundling increases the wider the slit is resulting
into complex (but marvellous looking) spider webs.
Further details on the search for vibrating suspended tubes
can be found in Ref.~\cite{us2}.
We argue that in the absence
of a support and at the relatively high temperature
CNTs may meet each other during growth.
The likelihood is increased if growth proceeds in `free' space
over a large distance. Once they touch each other they stick together
due to the van der Waals interaction leading to a bundle.

In contrast, the growth on a substrate is different, as the tubes
interact with the substrate rather than with each other. Hence,
bundling is expected to be reduced. This is confirmed in AFM images,
provided the catalyst density is low.
However, there are bundles as well, which is evident from the
observed branches visible in the AFM image of Fig.~1a (arrows).
Even at locations where bundling is not apparent, one can still not be sure that
such a nanotube section corresponds to a single-wall tube.
Usually this is checked by measuring the height in AFM, but this can be
misleading too, because the diameters of CVD-tubes can vary a lot, over
\mbox{$1-5$\,nm} as reported by Ref.~\cite{Dai2}. We confirm this with our
own measurements. Further insight into the question of bundling of
CVD-grown CNTs can be obtained from electrical
characterizations, which we report next.

\subsection{Carbon nanotube devices}

We have produced CNT devices on chip following two strategies. In
the first method the substrate is covered with a layer of
polymethylmethacrylate (PMMA) in which windows are patterned by
electron-beam lithography (EBL). Next, the catalyst is spread
from solution over these
patterned structures, after which the PMMA is
removed with acetone, leaving isolated catalyst islands
\mbox{($5\times 10$\,$\mu m^2$)} on the surface. The substrate
with the catalyst is then transferred to the oven where CVD growth
of CNTs is performed. From the catalyst islands, CNTs grow
randomly in all directions, but because of the relatively large
distance between the islands \mbox{(5\,$\mu m$)} just one
or a few CNTs bridge them usually.
An atomic force microscope (AFM) image
in phase mode with several CNTs growing from the catalyst islands
is shown in Fig.~2a. An individual SWNT bridging the catalyst
islands is shown in Fig.~2b.
Metal electrodes (Au, Ti, Pd) are patterned over the catalyst islands with EBL,
followed by evaporation and lift-off. The alignments during the
EBL structuring have been done corresponding to chromium markers
\cite{Chromium}. SEM and AFM images of contacted
individual CNTs are shown in Fig.~3a and 3b.

In the second method we spread the (diluted) catalyst over the
entire substrate at low concentration. The density is chosen such
that at least one CNT grows inside a window of size
\mbox{$10\times 10$\,$\mu$m$^2$}. After the CVD process a set of
recognizable metallic markers (Ti/Au bilayer) are patterned, again
by EBL, see Fig.~3c. Using AFM in tapping mode, a suitable CNT
with an apparent height of less than \mbox{$3$\,nm} is located
with respect to the markers. In the final lithography step,
electrodes to the selected CNT are patterned by lift-off.

\subsection{Room temperature characterization}

Once the samples are made, it is common practice to distinguish
semiconducting and metallic CNTs by the dependence of their
electrical conductance ($G$) on the gate voltage ($V_g$), measured
at room temperature (\mbox{$T \approx 300$\,K}). This, however,
cannot be considered as a proof that an individual SWNT has been
contacted, because it is not well understood how the
linear response conductance is altered if more than one
tube is contributing to electrical transport.
Even if measurements were performed on ropes of
SWNTs, the measured signatures agreed quite well with the behavior
expected for a SWNT \cite{Tans2,Bockrath2,NygardKondo}. This has
been attributed to a dominant electrode-CNT coupling to one
nanotube only. This scenario may be true in exceptional
cases, but one would expect that the majority of measurements
should display signatures that arise from the
presence of more than one tube.
We have recently observed Fano resonances which we attribute to
the interference of a SWNT which is strongly coupled to the
electrodes with other more weakly
coupled ones~\cite{us1}.

Assuming that all chiralities have equal probability to be formed in growth,
\mbox{2/3} of the SWNTs are expected to be semiconducting and
\mbox{1/3} metallic. From the measured response of the
electrical conductance to the gate voltage (back-gate), \mbox{$\approx 60$\,\%} of
the devices display metallic
(the conductance does not depend on the gate voltage) and
\mbox{$\approx 40$\,\%} semiconducting behavior.
Based on our assumption the larger fraction of metallic gate responses
points to the presence of bundles or multishell tubes. If there are
on average $2$ or $3$ tubes per bundle, which are coupled to the electrodes
approximately equally, the probability to observe a semiconducting
characteristic would amount to \mbox{$(2/3)^2=44$\,\%} or
\mbox{$(2/3)^3=30$\,\%}. Hence, we can conclude that the bundle size is
very likely small and close to $2$ on average.

A powerful method to characterize contacted CNTs is to perform
transport measurements in the nonlinear transport regime (high
bias). As previously reported by Yao \textit{et al.} \cite{Kane1}
the emission of zone-boundary or optical phonons is very effective
in CNTs at high fields. This effect leads to a saturation of the
current for an individual SWNT at \mbox{$\approx 25$\,$\mu$A}.
High bias \mbox{$I/V$} characteristics are shown in Fig.~4.
Fig.~4a corresponds to an individual SWNT. The saturation current
can be extracted from the relation for the electrical resistance
\mbox{$R \equiv V/I$} \cite{Kane1}
\begin{equation}\label{HB}
R=R_0+V/I_0,
\end{equation}
where \mbox{$R_0$} is a constant and \mbox{$I_0$} is the saturation current.
The dependence of $R(V)$ versus the bias voltage $V$ is shown in the insets of
Fig.~4 with corresponding fits to Eq.~\ref{HB}.
Because the saturation current is relatively well defined, its measurement allows to
deduce the number of participating CNTs. Whereas Fig.~4a corresponds to a single SWNT, two
nano\-tubes seem to participate in transport in the example shown in Fig.~4b.
This result is consistent with the one above and points to the presence of
more than one tube. This saturation-current method
works for SWNTs but also for multi-wall CNTs \cite{Avouris}.
One can therefore not distinguish whether one deals with two
tubes in a rope or with one double-wall CNT.

\section{Low-ohmic contacts}

It is well known that physical phenomena explored by
electrical transport measurements (especially at low temperatures)
dramatically depend on the transparency between the contacts and
the CNT. At low energies, the electronic transport through an
ideal metallic single-wall carbon nanotube (SWNT) is
governed by four modes (spin included). In the Landauer-B\"uttiker
formalism \cite{B} the conductance can be written as
\begin{equation}\label{CNT1}
G=T \cdot 4 e^2/h,
\end{equation}
where \mbox{$T$} is the total transmission probability
between source and drain contacts. For low transparent contacts
(\mbox{$T<<1$}) the CNT forms a quantum dot (QD)
which is weakly coupled to the leads. Charge transport is then
determined by the sequential tunnelling of single electrons (Coulomb blockade regime).
If the transmission probability is increased (for which better
contacts are required), higher-order
tunnelling processes (so called co-tunnelling) become important which
can lead to the appearance of the Kondo effect. This phenomenon was first
reported by Nyg{\aa}rd {\it et al.}~\cite{Kondo4}.
At transparencies approaching $T\approx 1$
we enter the regime of ballistic transport where residual backscattering at the
contacts leads to Fabry-Perot like resonances \cite{FB}.
Good contacts with transparencies close to one are indispensable
for the exploration of superconductivity \cite{Helen}, multiple Andreev
reflection \cite{Mark2} or spin injection \cite{Nygard1} in CNTs.
Nevertheless, modest progress has been made so far on the control
of the contact resistances between CNTs and metal leads. Annealing
is one possible route, as proposed by the IBM group \cite{Martel} and
we confirm their results here. We compare in the following
Ti, Au and Pd contacts.

\subsection{Comparison between Ti, Au and Pd contacts}

In the ideal case of fully transmissive contacts, a metallic SWNT is expected to have
a conductance of \mbox{$G=4e^2/h$} (two modes), which corresponds
to a two-terminal resistance of \mbox{$6.5$\,k$\Omega$}.
In case of contacts made by
\mbox{Ti}, \mbox{Ti/Au} or \mbox{Cr} on as grown SWNTs, most of the
devices show resistances in the range between
\mbox{$100$\,k$\Omega$} to \mbox{$1$\,M$\Omega$}. In contrast, \mbox{Au} contacts
are better, because the measured resistances range typically between
\mbox{40\,k$\Omega$} and \mbox{100\,k$\Omega$}. Even for the
highest conductive sample the transmission probability is rather small and
amounts to only \mbox{$T \approx 0.16$} (per channel).

To lower the contact resistances we added an annealing step to the process,
which was motivated by the work of R.~Martel \textit{et al.} \cite{Martel}.
We have performed annealing on more than
\mbox{50} samples in a vacuum chamber fitted with a heating stage
at a back-ground pressure of \mbox{$<10^{-5}mbar$}. The resistance
is first recorded on as prepared devices. Then, they are
annealed with temperature steps of \mbox{100\,$^\circ$C} for
\mbox{5\,min} starting at \mbox{500\,$^\circ$C}. The results for
titanium and gold contacts are shown in Fig.~5a and 5b,
respectively.

In agreement with previous work \cite{Martel}
we find a pronounced resistance decrease for Ti contacts, if annealed
at \mbox{800\,$^\circ$C}. It was suggested by
R. Martel \textit{et al.} \cite{Martel} that the origin of the
resistance decrease is the formation of titanium carbide
(\mbox{Ti$_x$C}) at temperatures over \mbox{700\,$^\circ$C}.
In contrast to Ti contacts, we do not observe a dramatic change in the sample resistance
versus annealing temperature in case of Au contacts.
This suggests that unlike Ti on carbon
no chemical reactions take place between Au and carbon even at
temperatures as large as \mbox{800\,$^\circ$C}. We have also compared
annealing in vacuum with annealing in hydrogen within the same temperature
window (not shown). The outcome in terms of resistance change
is comparable to the vacuum results provided that  \mbox{$T < 700$\,$^\circ$C}.
At temperatures above  \mbox{$\approx 700$\,$^\circ$C}
the majority of the devices display a short to the back-gate.
We think that the reducing atmosphere is very effective
in partially etching the \mbox{SiO$_2$} at these high
temperatures.

Finally, we have also studied as-grown Pd contacts, which
were recently reported to lead to contacts that are lower ohmic than Au \cite{Dai2}.
In our own work (Fig.~5c) we have indeed found
independently of Javey~\textit{et al.} \cite{Dai2}
that palladium makes excellent contacts to CNTs. There is no
need for an additional post-growth treatment
\cite{us1}. Metallization of CNT devices with Pd is the preferred
method, because it yields low-ohmic contacts without an additional
annealing step. Careful transport studies of Pd contacted SWNTs
show Coulomb blockade, Kondo physics and Fano resonances
\cite{us1}. The observed resonances suggest that even in
nano\-tubes, which look at first sight ideal, interference with
additional transport channels may appear. The only plausible
explanation for this observation is the existence of other tubes,
hence a bundle or multishell nanotube.

\section{Conclusion}

Many applications of carbon nano\-tubes (CNTs) require to reproducibly place and contact
single {\em small} diameter tubes. This is important, for example, for the realization of
mechanical resonators~\cite{us2}, for field-effect transistors with reproducible
characteristics and for fundamental studies of electron transport.
One approach is to start from a powder of CNTs which is obtained, for example, in arc-discharge
or laser-evaporation. Because these methods yield bundles of dozens of tubes, individual
CNTs can only be obtained by rigorous ultrasonics and separation in an ultracentrifuge
in the presence of a surfactant. If the ultrasonic step is too rigorous, the CNTs are
cut into short pieces. Spreading and contacting of single tubes is possible. However, one has to
bear in mind that these CNTs are covered by a surfactant which is likely to affect the
fabrication of low-ohmic contacts. Moreover, the surfactant may carry charge which dopes
the CNTs. In contrast to this approach, chemical vapor deposition (CVD) yields tubes
in a very direct way immediately on the chip and without a surfactant, which makes this
approach very attractive. Whereas a profound comparison of the quality in terms of
the number of defects between these two major classes of CNTs is not yet established, the degree
of bundling can be compared today. If grown by CVD on a surface at relatively high
temperature and with a low catalyst density, apparently single-wall CNTs can be grown, though
with a much larger spread in diameter as compared to e.g. the laser method. Although, the
tubes appear to be single, as judged from SEM and simple tapping-mode AFM in air, we find
in a number of different experiments clear signs for the presence of more than one tube.
Measured saturation currents are often larger than the value expected for a single tube.
Suspended tubes, even if no bundling is apparent in SEM in the form of branches, do not thermally
vibrate as expected for a typical SWNT \cite{us2}. And finally, the presence
of interference effects in transport (Fano resonances) point to additional transport
channels that are likely due to additional shells or tube \cite{us1}.
The results presented in this
work show however, that the number of tubes can be small, e.g. 2-3. This gives hope that
with refined catalysts, the controlled production of single tubes should be possible.
In addition, we have demonstrated that relatively low-ohmic contacts can
be achieved either with Ti, if an additional annealing step is used, or by Au and Pd
without any additional treatment. Out of these three materials, Pd yields the best
contacts (lowest contact resistance).

%%%%%%%%%%%%%%%%%%%%%%%%%%%%%%%%%%%%%%%%%%%%%%%%
%% BACKMATTER
%%%%%%%%%%%%%%%%%%%%%%%%%%%%%%%%%%%%%%%%%%%%%%%%

\begin{acknowledgments}
We acknowledge contributions and discussions to this work by
T.~Y.~Choi (ETHZ), J.~Gobrecht (PSI), and D.~Poulikakos (ETHZ).
Support by the Swiss National Science Foundation, the NCCR on Nanoscience and
the BBW is gratefully acknowledged.
\end{acknowledgments}

\begin{figure}[ht]
\caption{\label{Figure. 1} SEM images of CNTs grown from
\mbox{catalyst~1}. In (a) the CNTs were grown on a
\mbox{Si/SiO$_2$} substrate at \mbox{$T$=800\,$^\circ$C}. The
arrows point to visible branches. (b) Typical CNT network, grown
over structured slits at \mbox{$T$=750\,$^\circ$C}. Note, that
CNTs can bridge very large distances.}
\end{figure}

\begin{figure}[ht]
\caption{(a) Phase image recorded by tapping mode AFM, showing
CNTs grown from the patterned catalyst islands and bridging
between islands. (b) Topography image of an individual SWNT grown
between the catalyst islands recorded by tapping mode AFM. Inset:
Height measurement on the line cut (white line) for the SWNT shown
in (b). The height measurements yield for the diameter
\mbox{$d=(1.2 \pm 0.2$)\,nm} for this particular tube.}
\end{figure}

\begin{figure}[hb]
\caption{\label{Figure. 3} (a) SEM image of a SWNT contacted with
a \mbox{Ti/Au} bilayer. (b) AFM image recorded in tapping mode of
a contacted individual SWNT. (c) SEM image of a set of
\mbox{Ti/Au} markers which are used to register the contact
structure to the SWNTs selected before by AFM.}
\end{figure}

\begin{figure}[hb]
\caption{ Typical \mbox{$I-V$} characteristics at high bias
voltage for CNT samples with a contact spacing of
\mbox{1\,$\mu$m}. The insets show \mbox{$R \equiv V/I$} versus $V$
and fits to Eq.~\ref{HB} for positive and negative $V$ (lines).
(a) The extracted mean value for the saturation current for this
device is \mbox{$I_0=24.3 \pm 1.2$\,$ \mu$A} which suggests
transport through an individual SWNT. (b) A higher saturation
current of \mbox{$I_0=59.3 \pm 2.1$\,$ \mu$A} is found in this
device suggesting transport through $2-3$ CNT shells.}
\end{figure}

\begin{figure}
\caption{
  Comparison of the two-terminal resistance $R$ at room temperature
  of CNT devices which were contacted with different metals: (a) Ti, (b) Au and
  (c) Pd. Post-annealing has been done in vacuum (\mbox{$<10^{-5}mbar$}) in case
  of Ti and Au. In (a) and (b) the evolution of $R$  for a large number \mbox{($\approx 55$)}
  of samples as a function of annealing temperature is displayed in the form of
  a histogram. The representation for Pd (c) is different: the conductance $G=1/R$
  of $10$ samples are compared, out of which only one has a resistance \mbox{$R>50$\,k$\Omega$}, corresponding
  to \mbox{$G<0.5$\,$e^2/h$}.
}
\end{figure}


\begin{thebibliography}{99}
\bibitem{flowmeters}
 Calibrated massflow controllers were used, except for the large flow rate
 of methane of \mbox{$5000$\,cm$^3$/min}, which was adjusted and meseared
 with a floating ball meter calibrated to air. The correction factor for
 methan is $\approx 1.4$, i.e. the actual flow was $1.4$ times larger.
\bibitem{Chromium}
 Chromium is used because it could sustain high temperature of CVD
  process with no apparent diffusion.
\end{thebibliography}
\end{document}